\input{aipcheck}

\documentclass[
    ,final            
  ]
  {aipproc}

\layoutstyle{6x9}

\begin{document}

\newcommand{\lP}{\ell_{\rm P}}
\newcommand{\md}{{\mathrm d}}
\newcommand{\tr}{{\mathrm{tr}}}
\newcommand{\vt}{\vartheta}
\newcommand{\vp}{\varphi}

\title{Quantum geometry and quantum dynamics\\ at the Planck scale}


\author{Martin Bojowald}{
  address={Institute for Gravitation and the Cosmos, The Pennsylvania State University,\\ 104 Davey Lab, University Park, PA 16802, USA}
}

\begin{abstract}
  Canonical quantum gravity provides insights into the quantum
  dynamics as well as quantum geometry of space-time by its
  implications for constraints. Loop quantum gravity in particular
  requires specific corrections due to its quantization procedure,
  which also results in a discrete picture of space. The corresponding
  changes compared to the classical behavior can most easily be
  analyzed in isotropic models, but perturbations around them are more
  involved.  For one type of corrections, consistent equations have
  been found which shed light on the underlying space-time structure
  at the Planck scale: not just quantum dynamics but also the concept
  of space-time manifolds changes in quantum gravity. Effective line
  elements provide indications for possible relationships to other
  frameworks, such as non-commutative geometry.
\end{abstract}

\maketitle


\section{Space-time structure}

Canonical formulations provide insights in underlying symmetries,
which for gravity correspond to general covariance. Once quantized,
correction terms result which may change the underlying symmetries or
even provide new quantum degrees of freedom.  In loop quantum gravity,
corrections arise from quantum geometry (the spatial structure) as
well as quantum dynamics. The main recent developments to be described
here, obtained from model systems or perturbations, are (i) consistent
deformations of classical gravity, and (ii) effective descriptions to
derive interacting quantum states and quantum corrections in equations
of motion.

\subsection{Canonical gravity}

For gravity, we have an infinite dimensional phase space of fields
$q_{ab}$ (the spatial metric) and momenta $p^{ab}$ (related to
extrinsic curvature). The other components of the space-time
metric, lapse $N$ and shift $N^a$ in
\[
  \md s^2=g_{\mu\nu}\md x^{\mu}\md x^{\nu}=
-N^2\md t^2+q_{ab}(\md x^a+N^a\md t) (\md x^b+N^b\md t)\,,
\]
are not dynamical since $\dot{N}$ and $\dot{N}^a$ do not occur in the
action. They may be included in an extended phase space, but their
momenta $p_N=\delta S/\delta\dot{N}$ and $p_{N^a}= \delta S/\delta
\dot{N}^a$ would be constrained to vanish identically. Accordingly,
$\dot{p}_N=-\delta S/\delta N$ and $\dot{p}_{N^a}=-\delta S/\delta
N^a$ must vanish, too, which implies additional constraints
\[
 C=\frac{\sqrt{\det q}}{16\pi G}{}^{(3)}\!R- \frac{16\pi
G}{\sqrt{\det q}} (p_{ab}p^{ab}-{\textstyle\frac{1}{2}}(p^a_a)^2)=0
\quad,\quad C_a= 2D_bp_a^b=0
\]
for the non-trivial phase space variables.

In addition to constraining the fields and their initial values, the
constraints generate gauge transformations which must leave physical
observables unchanged. The diffeomorphism constraint $
D[N^a]=\int\md^3x N^aC_a$ generates spatial diffeomorphisms along a
vector field $N^a$, while the Hamiltonian constraint $H[N]= \int\md^3x
NC $ completes this to space-time transformations (for fields
satisfying the constraints).

The dynamics of general relativity in a canonical formulation is
determined completely by the constraints, forming a total constraint
$T[N,N^a]=\int\md^3x (NC+N^aC_a)=0$ for all multiplier functions $N$,
$N^a$.  It generates equations of motion $\dot{f}=\{f,T[N,N^a]\}$ for
any phase space function $f(q,p)$, a dot referring to the time gauge
as given by the choice of lapse $N$ and shift $N^a$ to be inserted in
$T[N,N^a]$.

An important consistency requirement follows from this setup: if the
constraints must vanish at all times, the time derivative
$\dot{T}[M,M^a]=\{T[M,M^a],T[N,N^a]\}=0$ must vanish for all $N$ and
$N^a$. The right part of this equation,
\begin{equation} \label{FirstClass}
 \{T[M,M^a],T[N,N^a]\}=0 \quad\mbox{if}\quad T[N,N^a]=0\,,
\end{equation}
implies that the constraints form a so-called first class algebra
under Poisson brackets. By general covariance, this is automatically
satisfied classically: the constraints of general relativity are
covariant under space-time diffeomorphisms. As an important
consistency condition, (\ref{FirstClass}) must be realized also after
quantum corrections have been included in $T[N,N^a]$. If it remains
satisfied, an anomaly-free version of quantum effects has been
achieved. Since this condition is very restrictive, the
anomaly-problem remains one of the most important issues in
(canonical) quantum gravity.

\begin{figure}
  \includegraphics[height=.13\textheight]{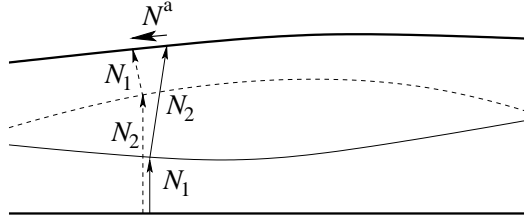}
  \caption{Illustration of the hypersurface deformation algebra.
    \label{f:SurfaceDef}}
\end{figure}

If we look at the constraint algebra for general phase space
configurations, not just for fields satisfying the constraints, more
information about space-time structure can be obtained. Working out
the specific constraint algebra for gravity, we obtain
\begin{equation}\label{AlgClass}
 \{H[N_1],H[N_2]\} = D\left[q^{a b}
 (N_1\partial_b N_2-N_2\partial_b N_1)\right]
\end{equation}
for the bracket of two Hamiltonian constraints (while brackets
involving $D[N^a]$ directly reflect the action of spatial
diffeomorphisms on $N$ or $N^a$). This clearly vanishes once the
diffeomorphism constraint is satisfied, but the structure for general
fields gives us a wider perspective on the types of gauge
transformations involved. If one looks at other generally covariant
systems, not just general relativity but also modified versions such
as those including extra fields or higher curvature terms, one finds
the same constraint algebra. The constraints certainly change, and so
does the dynamics, but they still satisfy the same algebra as shown
above. The algebra is thus very basic, depending not on the dynamics
of the theory but only on the space-time structure. In fact, one can
interpret it as generating transformations which correspond to
deformations of the hypersurfaces underlying the foliation used to set
up the canonical formalism. Accordingly, it is called hypersurface
deformation algebra, illustrated in Fig.~\ref{f:SurfaceDef}.

The advantage of using the constraints to generate space-time
transformations, rather than Lie derivatives along space-time vector
fields, is that this procedure directly deals with objects available
after quantization. Neither the manifold nor coordinates will be
related to operators, or be represented otherwise in the resulting
quantum theory.  Their form would have to come out much more
indirectly in canonical quantum gravity, but the constraints and their
algebra is one of the most basic and important aspects. Although
consistently implementing them is certainly not easy, they give us a
more direct handle on aspects of quantum space-time.

\subsection{Quantum corrections}

Canonical quantization following the Dirac procedure turns the
constraints into operators and requires them to annihilate physical
states. A well-defined quantization is often based on specific
constructions which lead to terms in the constraints not seen
classically. One, spatial discreteness, is a consequence of some
approaches to quantum gravity and leads to deviations from the
continuum expressions.  After quantization, the constraints typically
change, and so does their algebra. In this way, one can analyze
consequences for the quantum space-time structure without direct
reference to manifolds or coordinates, which would no longer be part
of the quantum theory.

Specifically, there are three types of corrections in loop quantum
gravity \cite{Rov,ALRev,ThomasRev}, which in general are equally
important:
\begin{itemize}
\item Entire states evolve which spread and
deform. Quantum fluctuations, correlations and higher moments
are independent variables back-reacting on expectation values. This is
well-known from quantum mechanics and quantum field theory, where its
effects can be captured by the loop expansion.
\item In loop quantum gravity, holonomies
\begin{equation}\label{Hol}
 h_e(A)={\cal P}\exp\left(\int_e A_a^i\tau_i\md t\right)
\end{equation}
as non-local, non-linear functions imply higher order corrections when
they appear in constraints in place of the connection $A_a^i$
\cite{LoopRep}.
\item Again in loop quantum gravity, fluxes 
\begin{equation}\label{Flux}
 F_S(E)= \int_S\md^2y E^a_in_a
\end{equation}
quantizing the spatial metric have discrete spectra containing zero.
Inverse metric components receive corrections for discrete
(lattice-like) states with small elementary areas \cite{QSDV}.
\end{itemize}
Canonical methods for the first class of effects are available
\cite{EffAc}, including the treatment of constraints
\cite{EffCons,EffConsRel} (but net yet fully extended to fields). The
latter two directly probe quantum geometry, intimately tied to the
discreteness.

\section{Quantum Friedmann equation}

Several of these effects can be illustrated easily
in isotropic cosmological models. In this case, they result in
a corrected Friedmann equation which can be summarized as
\begin{equation}
 \left(\frac{\dot{a}}{a}\right)^2 = \frac{8\pi G}{3}\left(\rho
 \left(1-\frac{\rho_Q}{\rho_{\rm crit}}\right)
 +\frac{1}{2}\sqrt{1-\frac{\rho_Q}{\rho_{\rm crit}}} 
\eta (\rho-P)+ \frac{(\rho-P)^2}{(\rho+P)^2}\eta^2
\right)
\end{equation}
where $P$ is pressure in addition to the energy density $\rho$
\cite{BounceSqueezed}; for earlier versions see
\cite{DiscCorr,GenericBounce,SemiClassEmerge,RSLoopDual}.  Quantum
effects thus make the Friedmann equation pressure dependent, which
would not appear in the classical equation. Moreover, there are
corrections depending on the form of the quantum state (of gravity,
not matter). First, $\eta$ parameterizes quantum correlations; it
would vanish only for a completely uncorrelated state. Secondly,
quantum fluctuations, or a whole series of fluctuation parameters
$\epsilon_k$ obtained from higher moments of the state, define
\[
 \rho_Q:=\rho+\epsilon_0 \rho_{\rm crit}+ (\rho-P) \sum_{k=0}^{\infty}
\epsilon_{k+1}
 (\rho-P)^k/(\rho+P)^k
\]
This corrected, quantum density appears in a term together with the
critical density $\rho_{\rm crit}=3/8\pi GL(a)^2$, which does not
introduce a new degree of freedom but depends on a characteristic
length scale $L(a)$ used in setting up the reduced model. (The size of
this parameter and its possible $a$-dependence are currently not
fixed, but should in principle be derived from a full theory once the
underlying state is under control. Till then, constraints on its value
can be found by internal consistency or phenomenology.)

In general, the dynamics is complicated since not only $a$ or the
matter field would be dynamical, but also the moments contained in
$\eta$ and $\rho_Q$: The entire state evolves, not just its
expectation values giving rise to $a$ and the classical matter
variables.  We are forced to deal with a higher-dimensional effective
system, containing new quantum degrees of freedom.  However, there are
cases in which the behavior simplifies. If $\rho=P$, for instance,
which is realized for a free, massless scalar, the state parameters
decouple: the model becomes exactly solvable and free of quantum
back-reaction \cite{BouncePert,BounceCohStates}.  Then, the only
correction to the Friedmann equation is to replace $\rho$ with
$\rho(1-\rho/\rho_{\rm crit})$, implying bouncing solutions
\cite{RSLoopDual}. This special model was in fact first studied in
considerable detail, mainly by numerically solving for the wave
function \cite{APS}. A similar equation results when $\eta=0$ around
the point where $\rho_Q=\rho_{\rm crit}$, but this seems to require
more special properties of the state.

The main origin of corrections giving rise to bouncing solutions is
the use of holonomies. By reformulating the higher order terms which
they initially imply for $\dot{a}$, they give rise to the quadratic
energy term in the Friedmann equation (see e.g.\ \cite{AmbigConstr}).
To understand all implications of the bounce for cosmology, including
metric perturbations at least at the linear level would be essential.
Unfortunately, so far no consistent inhomogeneous formulation with
holonomy corrections is available. Available equations either apply
only to special modes \cite{Vector,Tensor}, or are based on gauge
fixing \cite{HolonomyInfl,BounceCMB}. In the latter case, anomalies or
crucial quantum effects are hidden, but would have important
consequences for the behavior. (Examples are mentioned below.) It thus
remains unclear how bounces as they can sometimes be obtained from
effective Friedmann equations fit into a cosmological scenario.
Consistent equations for inhomogeneities in the presence of
corrections from loop quantum gravity are, however, available for
inverse metric corrections, to which we turn now. This brings us back
to the structure of space-time in canonical quantum gravity.

\section{Inverse metric corrections}

Inverse metric corrections change the Hamiltonian constraint, in its
gravitational part \cite{QSDI} as well as matter Hamiltonians
\cite{QSDV}, whenever inverse components of the densitized triad
appear. To avoid details and focus on generic implications, we can
implement them by a generic function $\alpha$ depending on the
gravitational phase space variables. In some cases, such as isotropic
models \cite{InvScale,Ambig} or regular lattice states with gauge
fixing \cite{QuantCorrPert} the form can be computed explicitly,
although it remains subject to quantization ambiguities; see
Fig.~\ref{f:alpha}.

\subsection{Number of spatial atoms}

Quantum geometry corrections of loop quantum gravity depend on the
form and size of discrete building blocks realized in the theory and
its states.  Geometrical operators have discrete spectra
\cite{AreaVol,Area,Vol2}, showing that the spatial geometry is made up
from small constituents. These constituents come in different sizes,
determined by the spin labels of a spin network state, and can form a
macroscopic geometry in many different ways.  Dynamically, one expects
the constituents to change in size as well as number, giving rise to the
evolution of a continuous geometry on large scales.

Specifically, this is realized in loop quantum cosmology by the
mathematical objects of holonomies (\ref{Hol}) and fluxes
(\ref{Flux}), whose elementary size in terms of coordinates we call
$\ell_0$ (linear for holonomies, quadratic for fluxes). In a region of
total coordinate volume $V_0$, the number of ``atoms of geometry'' is
then ${\cal N}=V_0/\ell_0^3$. This parameter depends on the size of
the region chosen, but not on coordinates. The density ${\cal
  N}/V_0=\ell_0^{-3}$, on the other hand, is independent of the region
but coordinate dependent. The only coordinate and region-independent
measure for the denseness of spatial atoms is the geometrical density
$\rho_{\cal N}= {\cal N}/a^3V_0= (\ell_0a)^{-3}$ in a universe of
scale factor $a$. Although $\ell_0$ is fixed for basic operators in
static quantum geometry, before the constraints are imposed, capturing
the full dynamics of changing lattices requires $\ell_0$ to depend on
``time'' \cite{InhomLattice}. This is to be understood in the internal
time sense, e.g.\ with reference to the scale factor: $\ell_0(a)$. The
same function determines the characteristic scale $L$ seen in the
critical density for Friedmann universes: $L(a)=a\ell_0(a)$.

These parameters enter basic holonomies via $\exp(i\ell_0\dot{a}/N)$
(with the lapse function $N$, which is one if the dot refers to proper
time) and fluxes via $\mu=\ell_0^2a^2/\ell_{\rm P}^2$ if the geometry
is nearly isotropic.  Strong quantum corrections (holonomies deviating
strongly from $\dot{a}/N$, or large inverse metric corrections
$\alpha(\mu)$) result in both cases if the arguments $\ell_0\dot{a}/N$
of holonomies or the values of fluxes are of the order one.  Holonomy
corrections thus appear when the curvature is $\dot{a}/N\sim k_*:=
({\cal N}/V_0)^{1/3}$.  Inverse metric corrections are large when
$a\sim a_*:= ({\cal N}/V_0)^{1/3}\ell_{\rm P}$. (Notice that these
equations are coordinate inpendent even though each side of the
inequalities depends on coordinates. For instance, for $a\sim a_*$,
the geometrical vertex density ${\cal N}/a^3V_0= (a_*/a)^3/\ell_{\rm
  P}^3$ is near one per Planck volume, a statement completely
independent of coordinates as well as $V_0$ or any region chosen.)
The classical range, when all corrections are small, is characterized
by $\dot{a}/N\ll k_*$ and $a\gg a_*$, as determined by the vertex
density ${\cal N}/V_0$ of a quantum geometry state.

For cosmological evolution, the dependence of ${\cal N}$ on $a$ is
important, which occurs whenever the discrete structure is being
refined during expansion: spatial atoms emerge dynamically, ensuring
that the discrete geometry is not enlarged to macroscopic sizes by
cosmic expansion. A derivation of this refinement from a full theory
of quantum gravity, including all inhomogeneities, is challenging. At
the current stage, this picture gives rise to different models
obtained by parameterizations.  If one assumes a power law ${\cal
  N}\propto a^{-6x}$, which can describe at least finite ranges of
evolution, one generically expects $-1/2<x<0$ according to what is
known about the dynamics of loop quantum gravity. The limiting cases
are interpreted as follows: For $x=0$, ${\cal N}$ is constant and
there is no refinement; discrete building blocks are just enlarged
during expansion. As one may expect, this leads to late-time problems
especially during the prolonged expansion of inflation
\cite{RefinementInflation,RefinementMatter}.  For $x=-1/2$, on the
other hand, we have a constant size of building blocks, and their
number increases proportionally to volume. There are no further
excitations of spatial ``atoms'' beyond their initial size.

Such effects, though schematic, have surprisingly strong consequences.
First, consistency bounds from the interplay of holonomy and inverse
metric corrections \cite{Consistent} or from more restrictive
anisotropic models \cite{SchwarzN} exist. Secondly, this allows us to
use phenomenology to see how quantum gravity dynamically refines its
discrete space. Recent examples have resulted, for instance, in an
upper bound ${\cal N}/a^3V_0<3/\ell_{\rm P}^3$ for the density from
big bang nucleosynthesis \cite{FermionBBN}, or a characteristic
blue-tilt for tensor modes which is enhanced if $x>-1/2$ while for
$x=-1/2$ correction are only small and of size $8\pi G\rho\ell_{\rm
  P}^2$ \cite{TensorHalf,TensorHalfII}.

\subsection{Linear perturbations}

Once a fully consistent theory for linear perturbations, including
inverse metric corrections, is set up and evaluated, restrictions on
quantum geometry might become even sharper.  For linear metric
perturbations around Friedmann--Robertson--Walker backgrounds (which
require the Hamiltonian to be expanded to second order) the corrected
Hamiltonian can be expanded as \cite{ConstraintAlgebra}
\[
H_{\rm grav}^Q := \frac{1}{16\pi G} \int\mathrm{d}^3x
\left(\bar{N}\!\left( \bar{\alpha}{\mathcal H}^{(0)} +
\alpha^{(2)}{\mathcal H}^{(0)} + 
\bar{\alpha}{\mathcal H}^{Q(2)}\right)+
 \delta N \bar{\alpha}{\mathcal H}^{Q(1)}\right)
\]
with ${\cal H}^{(0)}= -6{\cal H}^2a$ (the background Hamiltonian) and
\begin{eqnarray*}
{\mathcal H}^{Q(1)} &=& -4(1+f) {\cal H}a
\delta^c_j\delta K_c^j -(1+g)\frac{{\cal H}^2}{a}
\delta_c^j\delta E^c_j +\frac{2}{a}
\partial_c\partial^j\delta E^c_j  \\
{\mathcal H}^{Q(2)} &=& a \delta
K_c^j\delta K_d^k\delta^c_k\delta^d_j - a (\delta
K_c^j\delta^c_j)^2 -\frac{2{\cal H}}{a} \delta
E^c_j\delta K_c^j
\\
&& -\frac{{\cal H}^2}{2a^{3}} \delta E^c_j\delta
E^d_k\delta_c^k\delta_d^j
+\frac{{\cal H}^2}{4a^{3}}(\delta E^c_j\delta_c^j)^2
-(1+h)\frac{\delta^{jk} }{2a^{3}}(\partial_c\delta E^c_j)
(\partial_d\delta E^d_k)\,.
\end{eqnarray*}
Here, $\delta E^a_i$ and $\delta K_a^i$ are perturbations of the
densitized triad and extrinsic curvature, respectively.  In addition
to the primary correction function $\alpha$, with its background value
$\bar{\alpha}$ and the second order contribution $\alpha^{(2)}$, there
are extra function $f$, $g$ and $h$. They will be fixed in terms of
$\alpha$ later on.

\begin{figure}
  \includegraphics[height=.25\textheight]{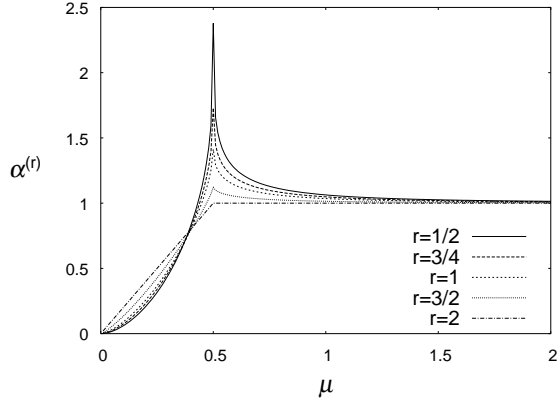}
  \caption{Examples for inverse metric correction functions, subject to 
 ambiguities parameterized by $0<r\leq 2$. \label{f:alpha}}
\end{figure}

Modifying the constraints in most cases leads to anomalies: classical
first class constraints then no longer form a first class algebra.
Severe consistency issues result, such as inconsistent equations or
the coupling of gauge parameters to observables. For a gauge system,
quantization or any form of quantum corrections of the constraints
must be a consistent deformation, respecting the first class
nature of the constraints. For constraints corrected by inverse triad
corrections as above, a first class algebra to second order in
perturbations is realized if the correction functions satisfy $2f+g=0$
and
\begin{eqnarray*}
 -h -f+\frac{a}{\bar{\alpha}}\frac{\partial
\bar{\alpha}} {\partial a} =0 \quad&,&\quad
f-g-2a\frac{\partial f}{\partial a}
-\frac{a}{\bar{\alpha}}\frac{\partial \bar{\alpha}}
{\partial a} =0 \\
\frac{1}{6}\frac{\partial\bar{\alpha}}{\partial a}
\frac{\delta E^c_j}{a^3}
+ \frac{\partial\alpha^{(2)}}{\partial(\delta E^a_i)}
(\delta^a_j \delta^c_i - \delta^c_j \delta^a_i) &=& 0\,.
\end{eqnarray*}
(There are additional conditions for matter correction functions in
terms of $\bar{\alpha}^2$ if matter is present.)  

With these equations, all initial coefficients are fixed in terms of
$\bar{\alpha}$, whose general form can be derived in models.
Importantly, corrections of inverse metric type are possible in a
consistent deformation: $\bar{\alpha}$ remains undetermined from the
algebra, and so need not take the classical value $\bar{\alpha}=1$. It
is allowed to be of the form seen in Fig.~\ref{f:alpha}, for instance.
Inverse metric corrections can be implemented in an anomaly-free form,
producing consistent equations for scalar linear perturbations
\cite{ScalarGaugeInv},
\[
\partial_c\left(\dot\Psi+{\cal H}(1+f)\Phi\right)=\pi
G\frac{\bar{\alpha}}{\bar{\nu}}\dot{\bar{\varphi}} \partial_c\delta\varphi^{\rm GI}
\]
from the diffeomorphism constraint,
\begin{eqnarray*}
&&\Delta(\bar{\alpha}^2
\Psi)-3{\cal H}(1+f)
\left(\dot\Psi+{\cal H}\Phi(1+f)\right)\\
&=&4\pi
G\frac{\bar{\alpha}}{\bar{\nu}}(1+f_3)
\left(\dot{\bar{\varphi}}\delta\dot\varphi^{\rm
GI}-\dot{\bar{\varphi}}^2(1+f_1)\Phi
+\bar{\nu} a^2 V_{,\varphi}(\bar{\varphi})
\delta\varphi^{\rm GI}\right)
\end{eqnarray*}
from the Hamiltonian constraint, and
\begin{eqnarray*}
&&\ddot\Psi+{\cal H}\left(2\dot\Psi\left(1-\frac{a}{2\bar{\alpha}}
\frac{{\rm d}\bar{\alpha}}{{\rm d}a}\right)+\dot\Phi(1+f)\right)
+\left(2\dot{\cal H}+{\cal H}^2\left(1+
\frac{a}{2}\frac{{\rm d}f}{{\rm d}a} -
\frac{a}{2\bar{\alpha}}\frac{{\rm d}\bar{\alpha}}{{\rm d}a}\right)\right)
\Phi(1+f)\\&&=4\pi G\frac{\bar{\alpha}}{\bar{\nu}}
\left(\dot{\bar{\varphi}}\delta\dot\varphi^{\rm
GI}-a^2\bar{\nu} V_{,\varphi}(\bar{\varphi})\delta\varphi^{\rm GI}\right)
\end{eqnarray*}
as the equation of motion. They show promising effects not appearing
classically, for instance the non-conservation of power on large
scales. Another implication is the existence of anisotropic stress, a
consequence not seen in gauge-fixed treatments: with the corrections,
$\Phi=(1+h)\Psi$.

\subsection{Quantum constraint algebra}

With the required conditions for $f$, $g$ and $h$ and $\alpha^{(2)}$,
the algebra of corrected constraints is first class: it presents a
consistent deformation of the classical theory to linear order in
inhomogeneities.  Anomaly-free constraints including quantum gravity
corrections thus exist. Even though the underlying discreteness, via
inverse metric corrections, is responsible for the occurrence of these
corrections, it does not destroy general covariance.

However, the constraint algebra of hypersurface deformations is
quantum corrected \cite{ConstraintAlgebra}:
\[
 \{H^Q[N_1],H^Q[N_2]\} = D\left[\bar{\alpha}^2\bar{N}a^{-1/2}
 \partial^a (\delta N_2-\delta N_1)\right]\,.
\]
(The same corrected algebra results in spherically symmetric models
without linearization \cite{SphSymmPSM,LTBII}.)  This may not be fully
surprising since the classical algebra (\ref{AlgClass}) contains the
inverse metric in its structure functions, and so inverse metric
corrections may be expected in the constraint algebra. What is
non-trivial is the conclusion that this correction can be implelented
anomaly-freely. The specific form obtained here indicates how the
structure of quantum space-time changes compared to the classical one:
even the constraint algebra, and thus the underlying algebra of
space-time diffeomorphisms, is corrected. Quantum gravity corrections
affect not only the dynamics of the theory, but also its underlying
symmetries. An immediate consequence in a canonical theory is that
quantum corrections to constraints change the form of gauge
invariant variables, as they appear in the consistent perturbation
equations presented before.  Here, differences to reduced phase space
quantizations arise, where classical gauge invariant quantities would
be quantized directly without implementing corrections to the gauge
behavior.

While it is clear that the quantum space-time structure must change
from inverse metric corrections, it is difficult to say what the new
manifold structure might be. We only know the constraint algebra so
far, which is difficult to integrate. Moreover, we only know the
corrections for linear perturbative inhomogeneities, and an extension
to higher orders or non-perturbative inhomogeneity is much more
involved. (But as suggested by non-linear spherically symmetric
models, it may well be possible.) An intriguing possibility, still to
be explored, would be a relationship to non-commutative geometry. In
both cases, canonical quantum gravity and non-commutative geometry,
manifolds are not taken as basic. But effective structures do arise,
which may be the best way to compare these different frameworks. If
such a relationship can be established, it might give indications for
deformed Lorentz symmetries in quantum gravity; see e.g.\ \cite{DSR}.

\section{Conclusions}

Different types of quantum corrections arise in loop quantum gravity:
those from quantum geometry (inverse metric/holonomy corrections) and
those from quantum dynamics (back-reaction). The former are specific
to the theory and thus provide useful opportunities for tests.

In particular the anomaly problem, which becomes severe in the context
of inhomogeneities and in particular with discreteness corrections,
can be addressed at an effective level. It turns out that consistent
deformations do exist, incorporating quantum effects from the inverse
metric (themselves coming from discrete flux spectra) in classical
equations. Via the consistent perturbation equations, an interface to
cosmological applications is obtained. Observational input is very
conceivable, and can shed light on the underlying quantum states by
constraining possible quantum corrections.

On a fundamental level, this tells us that discrete structures of
space-time do not have to break covariance. They may deform the
classical algebra, but the same number of symmetry generators remains
present. A different realization of covariance results, perhaps as a
deformed space-time diffeomorphism group.

\begin{theacknowledgments}
 Work reported here was supported in part by NSF grant PHY-0748336.
\end{theacknowledgments}



\bibliographystyle{aipproc}   


\IfFileExists{\jobname.bbl}{}
 {\typeout{}
  \typeout{******************************************}
  \typeout{** Please run "bibtex \jobname" to optain}
  \typeout{** the bibliography and then re-run LaTeX}
  \typeout{** twice to fix the references!}
  \typeout{******************************************}
  \typeout{}
 }

\end{document}